\newcommand{\anon}{1}

\documentclass[11pt]{article}

\usepackage{graphicx}
\usepackage{float}
\usepackage{lmodern}
\usepackage{booktabs}
\usepackage{tabularx}
\usepackage{amsmath,amsfonts,amssymb,bm,mathtools,relsize,exscale}
\usepackage{dsfont}
\usepackage{cancel}
\usepackage[OT1]{fontenc}
\usepackage[utf8]{inputenc}
\usepackage[linesnumbered,ruled,vlined]{algorithm2e}
\usepackage[inline,shortlabels]{enumitem}
\graphicspath{ {figs/} }

\usepackage{refcount,nameref,zref-xr,zref-user} % nameref before zref-xr
\zxrsetup{toltxlabel} % allows use of the \ref command as usual
\usepackage{url}

\usepackage{tikz}

\let\href\undefined
\usepackage[colorlinks,citecolor=blue,urlcolor=blue]{hyperref}

\makeatletter%
\@ifclassloaded{biometrika}{%
  \AtEndDocument{\printhistory}
  
  \def\rm{}

  \jname{Biometrika}
  \markboth{AUTHOR 1 ET AL.}{Great paper title}

  \newcommand{\authorlist}{
      \author{E.~TONG}
      \affil{Department of Statistics,
        Harvard College?,\\
        The department's address
      \email{auth1@g.harvard.edu}}

      \author{S.V.~BALKUS}
      \affil{Department of Biostatistics,
        Harvard Chan School of Public Health,\\
        665 Huntington Avenue, Boston, MA 02115
      \email{auth2@g.harvard.edu}}

  }
}{%
  \usepackage{arxiv}
  \usepackage{standalone}
  \usepackage{multirow}
  \usepackage{setspace}
  \usepackage{amsthm}
  \usepackage[round]{natbib}

  {\theoremstyle{definition}}
  {\theoremstyle{definition}}
  {\theoremstyle{definition}}

  \newcommand{\authorlist}{
    Eric Tong \\
    Department of Statistics,\\
    Harvard College,\\
    \texttt{erictong@college.harvard.edu}\\
    \And
    Salvador V.~Balkus \\
    Department of Biostatistics,\\
    Harvard Chan School of Public Health,\\
    \texttt{sbalkus@g.harvard.edu}\\
  }
}
\makeatother

\newcommand{\titlepaper}{Linear models for causal inference under network interference}

\newcommand{\makeabstract}{
In causal inference, interference occurs when the treatment of one unit may affect the outcomes of other units. The goal of this work is to serve as a guide to the use of linear outcome modeling for estimating causal effects in settings where interference may pose a challenge to identification and estimation, such as spatial and network data. We demonstrate that, under a linear model, causal effects of binary and continuous treatments can be identified in terms of regression coefficients under totally and partially known interference structures. Our work constructs unbiased and consistent point and variance estimators for these effects under one or more possible fixed or random interference networks. A chief advantage is that this approach can be implemented using standard linear regression software, and is easily augmented with random effects and heteroscedastic or autocorrelation consistent standard errors. Numerical experiments and an example data analysis demonstrate the efficacy of this approach in eliminating interference bias.
}

\AtEndDocument{\refstepcounter{theorem}\label{finalthm}}
\AtEndDocument{\refstepcounter{equation}\label{finaleq}}
\AtEndDocument{\refstepcounter{lemma}\label{finallemma}}

\newcommand{\Var}{\text{Var}}
\newcommand{\Cov}{\text{Cov}}

 \makeatletter
 \@ifclassloaded{biometrika}{%
   \authorlist
 }{
   \author{\authorlist}
   \date{\today}
}
\makeatother
\title{\titlepaper}

\begin{document}
\usetikzlibrary{arrows.meta, positioning}
\maketitle

%%% FOR TAS %%%
% \def\spacingset#1{\renewcommand{\baselinestretch}%
% {#1}\small\normalsize} \spacingset{1}

%%%%%%%%%%%%%%%%%%%%%%%%%%%%%%%%%%%%%%%%%%%%%%%%%%%%%%%%%%%%%%%%%%%%%%%%%%%%%%

% \if1\anon
% {
%   \title{\bf \titlepaper}
%   \author{Eric Tong\thanks{
%     The authors gratefully acknowledge the National Science Foundation (award no.~DGE 2140743) and the Harvard College and Harvard Data Science Initiative's SPUDS program for financially supporting this work.}\hspace{.2cm}\\
%     Department of Statistics, Harvard College\\
%     and \\
%     Salvador V. Balkus \\
%     Department of Biostatistics, Harvard T.H. Chan School of Public Health}
%   \maketitle
% } \fi

% \if0\anon
% {
%   \bigskip
%   \bigskip
%   \bigskip
%   \begin{center}
%     {\LARGE\bf \titlepaper}
% \end{center}
%   \medskip
% } \fi

% \bigskip
% \begin{abstract}
% \makeabstract
% \end{abstract}

% \noindent%
% {\it Keywords:} Outcome regression, spillover, random graph, treatment effects
% \vfill

% \newpage
% \spacingset{1.8} % DON'T change the spacing!

%%% arxiv abstract goes here
\begin{abstract}
    \makeabstract
\end{abstract}

\section{Introduction}

Linear regression is one of the main workhorses of modern-day statistics. One place it finds frequent application is in the field of causal inference. Under linearity and standard identification assumptions, estimating the regression coefficient for a treatment $A_i$ in a model for outcome $Y_i$ given confounders $L_i$ is equivalent to estimating the average treatment effect \citep{hernan2024}, a strategy known as \textit{outcome regression}. While far from the most sophisticated approach \citep{van2011targeted}, its ubiquity and familiarity renders it one of the simplest ways to estimate causal effects \citep{Lbke2020}. 

A key assumption underlying this strategy is that each unit's outcome $Y_i$ depends only on its \textit{own} treatment $A_i$ and confounders $L_i$, not those of other units. The violation of this assumption, when $Y_i$ may also depend on $A_j$ or $L_j$ for $j \neq i$, is called \textbf{interference}, or ``spillover'' \citep{Hudgens2008, Tchetgen2010}. When the interfering units are encoded in an adjacency matrix $G$, this is known as \textit{network interference} \citep{aronow_estimating_2017}. Figure~\ref{fig:dag} illustrates the causal structure. Interference commonly arises in spatial, network, and clustered data. For instance, in studies of air pollution and disease \citep{Wu2020} or social mobility \citep{Lee2024}, the population of one geographic unit may commute to other units, exposing individuals to pollution beyond their home geography. Ignoring this can cause the effect of air pollution on a desired outcome to be underestimated \citep{Shin2023}.

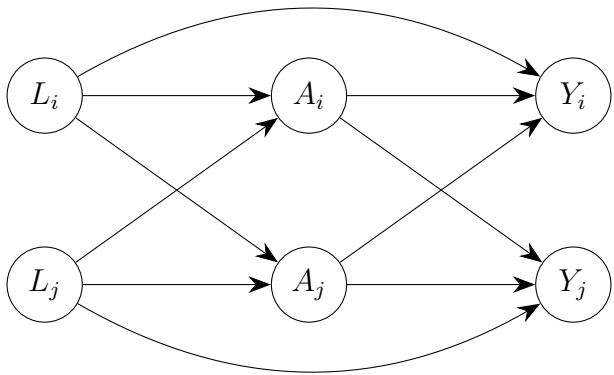
\begin{figure}[ht]
    \centering
    \begin{tikzpicture}[
    node distance=2cm and 2.5cm,
    mynode/.style={draw, circle, minimum size=1cm, font=\large},
    >={Stealth[length=3mm]}
]
    % Unit i nodes
    \node[mynode] (Li) {$L_i$};
    \node[mynode, right=of Li] (Ai) {$A_i$};
    \node[mynode, right=of Ai] (Yi) {$Y_i$};

    % Unit j nodes
    \node[mynode, below=1.5cm of Li] (Lj) {$L_j$};
    \node[mynode, right=of Lj] (Aj) {$A_j$};
    \node[mynode, right=of Aj] (Yj) {$Y_j$};

    % Local Causal Paths (Unit i)
    \draw[->] (Li) -- (Ai);
    \draw[->] (Li) to [bend left=30] (Yi);
    \draw[->] (Ai) -- (Yi);

    % Local Causal Paths (Unit j)
    \draw[->] (Lj) -- (Aj);
    \draw[->] (Lj) to [bend right=30] (Yj);
    \draw[->] (Aj) -- (Yj);

    % Interference (Spillover) Paths
    \draw[->] (Aj) -- (Yi);
    \draw[->] (Ai) -- (Yj);

    % Potential Confounding between units
    \draw[->] (Lj) -- (Ai);
    \draw[->] (Li) -- (Aj);
\end{tikzpicture}
\caption{Directed acyclic graph of causal relationships for ``interfering'' units $i$ and $j$.}
\end{figure}\label{fig:dag}

Long recognized as an obstacle to causal identification \citep{cox_planning_1958}, interference was first formalized as a component of the stable-unit treatment value assumption in causal inference by \cite{rubin1980}. Initial attempts to rectify the issue focused on design-based strategies and propensity score methods \citep{kempton1997interference, sobel2006randomized, hong2006evaluating}. \cite{Hudgens2008} and \cite{Tchetgen2010} introduced estimands decomposing the total effect into within-unit (``direct'') and spillover (``indirect'') effects under the assumption of \textit{partial interference}: that units can be partitioned such that only units within the same partition interfere with each other. Later, \cite{aronow_estimating_2017} formalized the notion of \textit{exposure mappings}, dimension-reducing summaries of neighboring treatments, in order to handle \textit{network} (or ``general'') interference, in which units may be interfere with each other according to their ties in a network. 

These dimension-reducing summaries lent themselves natural to model-based approaches, including machine-learning-based semiparametric methods for observational spatial data introduced by  \cite{vanderlaanCausalInferencePopulation2014} and further developed by 
\cite{sofryginSemiParametricEstimationInference2017, Ogburn2022, zivichTargetedMaximumLikelihood2022, Emmenegger2025} and \cite{Balkus2025}. Modern competing approaches decompose the joint treatment probability via the network structure \citep{tchetgentchetgenAutoGComputationCausalEffects2021, tchetgen2025autodrestimation, clarkCausalInferenceStochastic2024, Forastiere2020, Fritz2024}, or employ experimental and quasi-experimental approaches \citep{Aronow2012, Sinclair2012, Baird2018, Jiang2022, VazquezBare2023, Xu2023, Torrione2024}; see \cite{Bhadra2025} for a recent review.

Despite this rich literature, many existing methods pose a challenge to use in practice because they require bespoke estimation strategies or designs unavailable or unfamiliar to practitioners who may be more accustomed to standard regression workflows. For example, in spatial and clustered settings, analysts are often more familiar with linear mixed-effects models or generalized least squares \citep{gelman2007data}, or in econometrics, the spatial Durbin model \citep{anselin1988spatial}. Though such models can account for spatial autocorrelation or adjust for neighboring covariates, neither directly targets the effect of a causal intervention on neighbors' treatments. A couple works attempt to remedy this. \cite{LeSage2010} measure spillover via partial derivatives of spatial linear models, but connections to modern causal frameworks remains unclear, while \cite{VerbitskySavitz2012} use a hierarchical linear model to target treatment effects under spillover in an observational study of community policing. Our goal is to build on the work of \cite{VerbitskySavitz2012} to provide general, software-friendly tools for causal inference with linear models under interference.

\textbf{Contributions.} The purpose of this manuscript is to present a framework for linear outcome modeling in the presence of network interference. We argue that, under a linear model, treatment effect identification under interference reduces to an issue of model specification. Consequently, one can account for interference easily and naturally via slight modifications to estimation routines in existing statistical software like \texttt{R} \citep{r}. To this end, our work derives unbiased and consistent point and variance estimators for two common treatment effects of interest under totally and partially known interference structures, where the interference network may be fixed or random. Our results extend easily to complex cases where multiple possible interference networks may be considered. This work will guide general practitioners in using linear models to estimate causal effects in settings like spatial data where interference often lurks.

\textbf{Outline}. After a brief problem set-up (Section~\ref{sec:setup}), we present identification results and estimation methods (Sections~\ref{sec:identification}--\ref{sec:multiple-networks}), numerical simulations (Section~\ref{sec:simulation}), a real-data illustration (Section~\ref{sec:data-analysis}), and a discussion of limitations and future directions (Section~\ref{sec:discussion}).

\section{Methods: Linear models with interference}\label{sec:methods}

\subsection{Data and Problem Set-up}\label{sec:setup}

Consider $n$ exchangeable units $(L_i, A_i, Y_i)$, $i = 1, \ldots, n$, where $L_i$ is a vector of baseline covariates, $A_i$ a treatment or exposure, and $Y_i$ an outcome. By ``exchangeable'', we mean the joint distribution of the data is invariant to permutation of indices. We denote $Y = [Y_i]_{i=1}^n$ as the outcome vector, with analogous notation for $A$ and $L$. The data are supplemented by an adjacency matrix $G$, where $G_{ij} = 0$ indicates no interference between $i$ and $j$, and $G_{ij} \neq 0$ indicates that the treatment of unit $i$ may affect the outcome of unit $j$. $G$ may be binary or weighted (in which case $G_{ij}$ denotes the magnitude of interference), as well as directed or undirected. We define the \textit{neighborhood} of unit $i$ as $\mathcal{N}(i) = \{j : G_{ij} \neq 0\}$, and write $A_{\mathcal{N}(i)}$ and $L_{\mathcal{N}(i)}$ for the subvectors of treatments and covariates of $i$'s neighbors.

Assume the outcome follows the main-terms linear model
\begin{equation}\label{eq:original-model}
    Y_i = \beta_0 + \beta_a A_i + \sum_{j \in \mathcal{N}(i)} \beta_{a^s, j}G_{ij}A_j + \beta_lL_i + \sum_{j \in \mathcal{N}(i)}\beta_{l^s, j} G_{ij}L_j + \varepsilon_i
\end{equation}
where $E(\varepsilon) = 0$ and $\text{Var}(\varepsilon) = \sigma^2I_n + \rho S$ for some covariance matrix $S$ capturing spatial or network correlation; when $\rho = 0$ the errors are i.i.d. This is a classical linear model augmented by the weighted treatments and covariates of each unit's neighbors.

\subsection{Identification}\label{sec:identification}

This work will examine two estimands. For binary $A$, the \textit{average treatment effect} (ATE) is $E[Y(1) - Y(0)]$ \citep{hernan2024}; for continuous or count $A$, the 1-unit \textit{modified treatment effect} (MTE) is $E[Y(A+1) - Y(A)]$, also known as the population intervention effect of a 1-unit additive modified treatment policy \citep{munozPopulationInterventionCausal2012, haneuseEstimationEffectInterventions2013}. Under standard causal assumptions and the standard linear model $Y_i = \beta_a A + \beta_l L + \varepsilon_i$, these can both be identified as equal to the regression coefficient $\beta_a$. 

However, when interference is present as in Equation~\eqref{eq:original-model}, $\beta_a$ alone no longer identifies either estimand: $Y_i$ may be affected by an intervention that modifies its neighbors' treatments even if its own is unchanged. This violates causal consistency. Interestingly, this can also be seen as a form of model misspecification, as the terms of Equation~\eqref{eq:original-model} indexed by coefficients $\beta_{a^s,j}$ and $\beta_{l^s, j}$ have been excluded from the model (and thus, the identification formula). 

When the outcome $Y_i$ depends on both $i$'s own treatment and its neighbors', the potential outcomes become $Y_i(a_i, a_{\mathcal{N}(i)})$. In this case, we can attempt identification using natural extensions of the standard causal assumptions:
\begin{itemize}
    \item[\textbf{A1}] \textbf{(Consistency)} If $A_i = a_i$ and $A_{\mathcal{N}(i)} = a_{\mathcal{N}(i)}$, then $Y_i(a_i, a_{\mathcal{N}(i)}) = Y_i$.
    \item[\textbf{A2}] \textbf{(No unmeasured confounding)} $Y_i(a_i, a_{\mathcal{N}(i)}) \perp\!\!\!\perp (A_i, A_{\mathcal{N}(i)}) \mid (L_i, L_{\mathcal{N}(i)})$.
    \item[\textbf{A3}] \textbf{(Positivity for ATE)} $P(A_i = a_i, A_{\mathcal{N}(i)} = a_{\mathcal{N}(i)} \mid L_i = l_i, L_{\mathcal{N}(i)} = l_{\mathcal{N}(i)}) > 0$ for all $(a_i, l_i, a_{\mathcal{N}(i)}, l_{\mathcal{N}(i)})$ in the support.
    \item[\textbf{A4}] \textbf{(Positivity for MTE)} $(a_i+1, l_i, a_{\mathcal{N}(i)}+1, l_{\mathcal{N}(i)}) \in \text{support}(A, L)$ for all $(a_i, l_i, a_{\mathcal{N}(i)}, l_{\mathcal{N}(i)}) \in \text{support}(A, L)$.
\end{itemize}

The dependence of $Y_i$ on potentially many neighboring covariates can pose a challenge due to the high-dimensionality of the parameters. Fortunately, the assumptions of linearity and exchangeability together reduce the problem to a low-dimensional one. Permutation-invariance of the joint distribution implies that all neighbor-specific coefficients must be equal; formally $(\beta_{a^s,j}, \beta_{l^s,j}) = (\beta_{a^s}, \beta_{l^s})$ for all $j \in \mathcal{N}(i)$, otherwise exchangeability would not hold.  Hence, Equation~\eqref{eq:original-model} collapses to
\begin{equation}\label{eq:simple-model}
    Y_i = \beta_0 + \beta_a A_i + \beta_{a^s}A_i^s + \beta_lL_i + \beta_{l^s}L_i^s + \varepsilon_i
\end{equation}
where $A_i^s = \sum_{j \in \mathcal{N}(i)} G_{ij}A_j$ and $L_i^s = \sum_{j \in \mathcal{N}(i)} G_{ij}L_j$. In the language of \cite{aronow_estimating_2017}, \textit{linearity automatically enforces a known exposure mapping}. $Y_i$ can be thought of as a linear model of ``summary'' functions $A^s$ and $L^s$ that represent weighted sums of interfering neighbors.

With the parameter space reduced, identification in terms of regression coefficients can proceed. Let $F_i = \sum_{j \in \mathcal{N}(i)} G_{ij}$ denote the weighted degree of unit $i$. By linearity of expectation,
\begin{align*}
    E(Y_i \mid A_i = 1, L_i, A_{\mathcal{N}(i)} = \mathbf{1}) = \beta_0 + \beta_a + \beta_{a^s}F_i + \beta_l L_{i} + \beta_{l^s} L_i^s
\end{align*}
Subtracting the analogous expression at $A_i = 0$ and averaging over units gives
\[\psi_{\text{ATE}} = E(\beta_a + \beta_{a^s} F_i) = \boxed{\beta_a + \beta_{a^s}E(F_i)}\]
with an identical result for the MTE. Thus $\beta_a$ is the \textit{within-unit effect}---the direct impact of a unit's own treatment---while $\beta_{a^s}E(F_i)$ is the \textit{spillover effect} from neighbors, which grows with network density. Ignoring interference yields a total-effect bias of $\beta_{a^s}E(F_i)$ even if $\beta_a$ is estimated correctly. We refer to $\psi_{\text{ATE}}$ and $\psi_{\text{MTE}}$ (henceforth, just $\psi$) collectively as the ``total effect'', and to $\beta_a$ and $\beta_{a^s}$ as the within-unit and spillover effects.

\subsection{Estimation under totally known interference structures}\label{sec:totally-known}

In this section, we consider the case where a sample of $G$ is fully available to the investigator. Thus, treatment and confounder summaries are easily calculated by performing matrix-vector or matrix-matrix multiplications $GA$ and $GL$. We present how to estimate the ATE and MTE using these quantities. 

\textbf{Point estimation}. Our previously identified treatment effects can be estimated using the following procedure:
\begin{enumerate}
    \item Compute $GA$ and $GL$ and construct the augmented design matrix $\begin{bmatrix}A & L & GA & GL\end{bmatrix}$.
    \item Fit $E(Y \mid A, L) = \beta_a A + \beta_{l}L + \beta_{a^s}GA + \beta_{l^s}GL$ using standard software (e.g.\ \texttt{lm} in \texttt{R} \citep{r}) to obtain $\hat\beta_a$ and $\hat\beta_{a^s}$.
    \item Compute the empirical mean weighted degree $\bar{F} = \frac{1}{n}\sum_{i=1}^n F_i$.
    \item Form the plug-in estimate $\hat{\psi} = \hat{\beta}_a + \hat{\beta}_{a^s}\bar{F}$.
\end{enumerate}
Since $\hat{\beta}_a$, $\hat\beta_{a^s}$, and $\bar{F}$ are all unbiased conditional on $G$, by linearity of expectation and the law of total expectation,
\begin{equation}\label{eq:unbiased}
    E(\hat{\psi}) = E\Big[E(\hat{\beta}_a \mid G) + \bar{F}E(\hat{\beta}_{a^s} \mid G)\Big] = \beta_a + \beta_{a^s}E(F_i)
\end{equation}
and consistency follows by the continuous mapping theorem.

\textbf{Statistical inference}. Conditioning on the observed $G$ (so $\bar{F}$ is fixed), the variance of $\hat\psi$ can be derived by applying Bienayme's identity to obtain the following:
\begin{align}\label{eq:variance}
\Var(\hat {\psi} \mid G) = \boxed{ \Var(\hat \beta_a \mid G) + \bar{F}^2 \Var(\hat \beta_{a^s} \mid G) + 2\bar{F} \Cov(\hat \beta_a, \hat \beta_{a^s} \mid G) }
\end{align}

Estimates of $\Var(\hat\beta_a \mid G)$, $\Var(\hat\beta_{a^s} \mid G)$, and $\Cov(\hat\beta_a, \hat\beta_{a^s} \mid G)$ are returned by standard regression software (e.g.\ via \texttt{vcov()} in \texttt{R}) and can be plugged directly into Equation~\eqref{eq:variance}. Unbiasedness and consistency of this plug-in estimate follows if its components are each estimated unbiasedly and consistently. Wald-style $(1-\alpha)$ confidence intervals follow from the standard construction $\hat{\psi} \pm \Phi^{-1}(\alpha/2)\sqrt{\widehat{\Var}(\hat\psi \mid G)}$.

Even though typically only a single observation of $G$ is observed, in some cases it may be sensible to treat $G$ as random. In this case, the plug-in variance estimator is anti-conservative but can still achieve consistency. To show this, we can first apply the law of total variance:
\begin{align*}
   E(\widehat\Var(\hat{\psi} | G)) 
    &= E\Big[\widehat\Var(\hat \beta_a \mid G) + \widehat\Var(\bar{F}\hat \beta_{a^s} \mid G) + 2\widehat\Cov(\hat \beta_a,  \bar{F} \hat \beta_{a^s} \mid G)\Big] \\
    =& E\Big(E\Big[\widehat\Var(\hat \beta_a \mid G) \mid G\Big] + E\Big[\widehat\Var(\bar{F}\hat \beta_{a^s} \mid G) \mid G\Big] + 2E\Big[
    \widehat\Cov(\hat \beta_a, \bar{F} \hat \beta_{a^s} \mid G) \mid G\Big]\Big) \\
    =& E\Big[\Var(\hat \beta_a \mid G) + \Var(\bar{F}\hat \beta_{a^s} \mid G) + 2\Cov(\hat \beta_a, \bar{F}\hat\beta_{a^s}\mid G)\Big)\Big]\\
    =& \Big(\Var(\hat \beta_a) - \Var(E(\hat \beta_a \mid G)) \Big) + \Big(\Var(\bar{F}\hat \beta_{a^s}) - \Var[E(\bar{F}\hat \beta_{a^s}\mid G)]\Big)\\ &+ 2\Big(\Cov(\hat \beta_a, \bar{F}\hat\beta_{a^s}) - \Cov[E(\hat \beta_a \mid G), E(\bar{F}\hat\beta_{a^s} \mid G)] \Big)
\end{align*}
Recall $E(\hat\beta_a \mid G) = \beta_a$; since the variance or covariance of a constant is 0, the first and third subtrahends are equal to 0. Then, since $E(\hat\beta_{a^s} \mid G) = \beta_{a^s}$, by variance properties and the definition of $\bar{F}$, $\Var(E(\bar{F}\hat \beta_{a^s}\mid G))$ simplifies so that that 
\begin{equation}\label{eq:variance-bias}
    E(\widehat\Var(\hat{\psi} | G))= \boxed{\Var(\hat{\psi}) - \frac{\beta_{a^s}^2}{n^2}\Var(W)}
\end{equation}
where $W = \sum_{i,j} G_{ij}$ is the total edge weight of $G$. When $W$ is fixed, as in the models of \cite{albert2002statistical} and \cite{watts1998collective}, where each node has a fixed degree, or when the weights of the network are normalized to sum to a particular value, $\Var(W) = 0$, yielding unbiasedness. But even if $W$ may vary, consistency can still be achieved provided its variance does not grow too rapidly. For example, in an Erd\H{o}s-R\'enyi graph with edge probability $p$, $\Var(W) = n(n-1)p(1-p)$; the bias is $O(1)$ for fixed $p$ but shrinks to $o(1/n)$ when $p = o(1/n)$, yielding asymptotically valid inference.

\textbf{Handling correlated errors}. In general, a fixed-effect linear regression estimate for $\hat\beta$ will always be unbiased under correct specification. However, unless $\varepsilon$'s entries are i.i.d. (unlikely in the types of spatial or network settings where interference might arise) obtaining unbiased and consistent estimates of the components of $\Var(\hat{\psi})$ will require accounting for possible heteroscedasticity or autocorrelation in $\varepsilon$.

Since all standard linear model theory applies to Equation~\eqref{eq:simple-model}, correlated errors pose no special difficulty. If $\varepsilon$ exhibits spatial autocorrelation, clustering, or repeated measures correlation, one can employ general least squares (GLS) or a linear mixed-effects model (LMM) with an appropriate random effect structure---including, in the undirected network setting, a covariance of the form $\hat{\sigma}^2I + \hat\rho G$ to model network-correlated errors. Under a correctly-specified correlation structure, necessary plug-in components of $\widehat{\Var}(\hat\beta)$ can be extracted from any standard GLS or LMM fit (e.g.\ via \texttt{vcov()} in \texttt{R}) and plugged directly into Equation~\eqref{eq:variance} to obtain valid inference for the total effect. Note, however, that in longitudinal settings these adjustments do not correct for time-varying confounding, which requires further structural assumptions \citep[see also Section~\ref{sec:multiple-networks}]{Robins1986, hernan2024}.

\subsection{Estimation under partially known interference structures}\label{sec:partially-known}

When the full graph $G$ is unavailable but the weighted degree $F_i = \sum_{j \in \mathcal{N}(i)} G_{ij}$ is observed for each unit, causal effects can still be identified under two additional assumptions:
\begin{itemize}
    \item[\textbf{A5}] \textbf{(Treatment independence)} $A_i \perp\!\!\!\perp L_{\mathcal{N}(i)}$ for all $i$; i.e.\ treatment is independent of neighbors' covariates.
    \item[\textbf{A6}] \textbf{(Equal effects)} $\beta_a = \beta_{a^s}$; the within-unit and spillover coefficients are equal.
\end{itemize}
A5 can hold under randomized designs or when treatment assignment does not depend on network position, ensuring $L_{\mathcal{N}(i)}$ does not confound estimation of $\beta_a$. A6 may be appropriate when treatment has the same magnitude of effect regardless of network location. This might be appropriate in a study of, for instance, air pollution, where a toxin might plausibly cause the same harm no matter its physical location of exposure. However, it may be inappropriate for a treatment such as a vaccine, where an individual's own treatment strengthens their immune system but their neighbor's treatment only reduces the probability of transmission. Under A5--A6, the total effect simplifies to
\begin{equation}\label{eq:total-effect-same}
    \psi = \beta_a\Big[1 + E(F_i)\Big]
\end{equation}
and $\beta_a$ can be estimated by fitting regressions with either of the following formulae:
\begin{align}
    Y &\sim A + L + F \tag{Fixed Effects}\\
    Y &\sim A + L + (1\mid F) \tag{Mixed Effects}
\end{align}
The plug-in estimator $\hat\psi = \hat\beta_a(1 + \bar{F})$ is unbiased by the same arguments as Equation~\eqref{eq:unbiased}, and its variance can be estimated by the plug-in estimator
\begin{equation}
    \widehat{\Var}(\hat{\psi} \mid G) = (1 + \bar{F})^2 \widehat{\Var}(\hat{\beta}_a \mid G)
\end{equation}
When $G$ is random, the bias in this variance estimator mirrors Equation~\eqref{eq:variance-bias}: it is zero for networks with a fixed sum of edges, and is asymptotically negligible for sparse graphs with $\Var(W) \lesssim  n^2$.

\subsection{Models that adapt to multiple possible interference structures}\label{sec:multiple-networks}

The additive structure of linear models naturally accommodates interference from multiple sources. Given networks $G_1, \ldots, G_K$ representing distinct interference mechanisms, the outcome model becomes
\begin{equation}\label{eq:multi-network}
    Y = \beta_0 + \beta_aA + \beta_lL + \sum_{k=1}^K \beta_{a_k^s}G_kA + \sum_{k=1}^K \beta_{l_k^s}G_kL + \varepsilon
\end{equation}
and the total effect is identified as $\psi = \beta_a + \sum_{k=1}^K\beta_{a_k^s}E(F_{k,i})$, where $F_{k,i} = \sum_{j \in \mathcal{N}_k(i)} G_{k,ij}$ is the weighted degree of unit $i$ in network $G_k$. As in Section~\ref{sec:totally-known}, plug-in estimates of this parameter with components obtained from fitting the linear regression in Equation~\eqref{eq:multi-network} will be unbiased and consistent. The variance of this estimator will be
\begin{equation*}
    \Var(\hat\psi \mid G_1,\ldots,G_K) = \Var(\hat \beta_a) + \sum_{k=1}^K\bar{F}_k^2 \Var(\hat \beta_{a_k^s}) + 2\sum_{k=1}^K\bar{F}_k \Cov(\hat \beta_{a}, \hat \beta_{a_k^s}) + 2\sum_{k<m}\bar{F}_k\bar{F}_m \Cov(\hat \beta_{a_k^s}, \hat \beta_{a_m^s})
\end{equation*}
which has an unbiased plug-in estimate for fixed networks. For random networks, by the same arguments as Section~\ref{sec:totally-known}, the plug-in variance estimator satisfies
\begin{equation}
    E(\widehat{\Var}(\hat{\psi})) = \Var(\hat\psi) - \frac{1}{n^2}\sum_{k=1}^K\beta_{a_k^s}^2\Var(W_k) - \frac{2}{n^2}\sum_{k < m} \beta_{a_k^s}\beta_{a_m^s} \Cov(W_k, W_m)
\end{equation}
where $W_k$ is the total edge weight of $G_k$. Since $\Cov(W_k, W_m) \leq \sqrt{\Var(W_k)\Var(W_m)}$, a plug-in variance estimate is (asymptotically) unbiased provided $\Var(W_k)$ grows slowly. 

We briefly highlight three useful applications of the multi-network framework.

\textbf{Higher-order neighbors}. If outcomes depend on ``neighbors' neighbors'' in addition to immediate neighbors, one can include $G^k$ (the $k$th matrix power of $G$) as an additional network in Equation~\eqref{eq:multi-network} to capture interference froom $k$th-order neighbors. All preceding theory applies, though care is needed when treating $G^k$ as random: the variance of its total edge count may grow faster than that of $G$ itself, depending on the network structure.

\textbf{Time-varying interference}. When interference evolves over time, one can incorporate a sequence of networks $G_1, \ldots, G_T$, one per time period, into a wide-format model summarizing each time period's covariates by the corresponding $G_t$. For causal inference under treatment-confounder feedback, such a model must be combined with parametric g-computation \citep{Naimi2016, hernan2024}; the main modification is that treatments and confounders at each time point are summarized by the time-varying $G_t$.

\textbf{Network uncertainty and model selection}. When the correct interference structure is uncertain, standard model selection tools---$R^2$, AIC \citep{Akaike1973}, cross-validated MSE---apply directly, since the model in Equation~\eqref{eq:multi-network} is an ordinary linear regression conditional on $G$. One can fit separate models for each candidate network and select the model achieving the best criteria; though, one must bear in mind that metrics like cross-validation may require modification to estimate  in the presence of correlated errors \citep{Rabinowicz2020}. One could also include summaries from each network within a single model and apply Lasso penalization \citep{Tibshirani1996} to select relevant terms, applying corrections to perform post-selection inference \citep{Lee2016}. 

\section{Simulation Experiments}\label{sec:simulation}

We evaluate the estimators from Section~\ref{sec:methods} via simulation in \texttt{R} \citep{r}, drawing from the data-generating process
\begin{align*}
    & L_1 \sim \text{Gamma}(3, 1), \quad L_2 \sim \text{Pois}(1), \quad L_3 \sim \text{Beta}(2, 5), \quad L_4 \sim \text{Bern}(0.6) \\
    & A \sim L_1 + 2L_2 + 3L_3 + 4L_4 + \varepsilon_A, \qquad Y \sim A + L_1 + 2L_2 + 3L_3 + 4L_4 + GA + \varepsilon_Y
\end{align*}
with $\varepsilon_A, \varepsilon_Y \sim \mathcal{N}(0, \Sigma)$. We ran two experiments: a homoscedastic setting ($\Sigma = I_n$) and a correlated setting ($\Sigma = 3I_n + 1.5G$). Interference graphs were generated using \textit{igraph} \citep{igraph}: Erd\H{o}s-R\'enyi (\texttt{sample\_gnp}, $p = 0.01$), Barab\'asi-Albert (\texttt{sample\_pa}, power $= 0.05$), and Watts-Strogatz (\texttt{sample\_smallworld}, dim $= 1$, nei $= 10$, $p = 0.05$). The target parameter is $\text{MTE} = 1 + n^{-1}\sum_i F_i$. Each combination of graph type and error structure was evaluated at $n \in \{100, 400, 900, 1600\}$ over 100 replicates.

\begin{figure}[ht]
    \centering
    \includegraphics[width=0.75\textwidth]{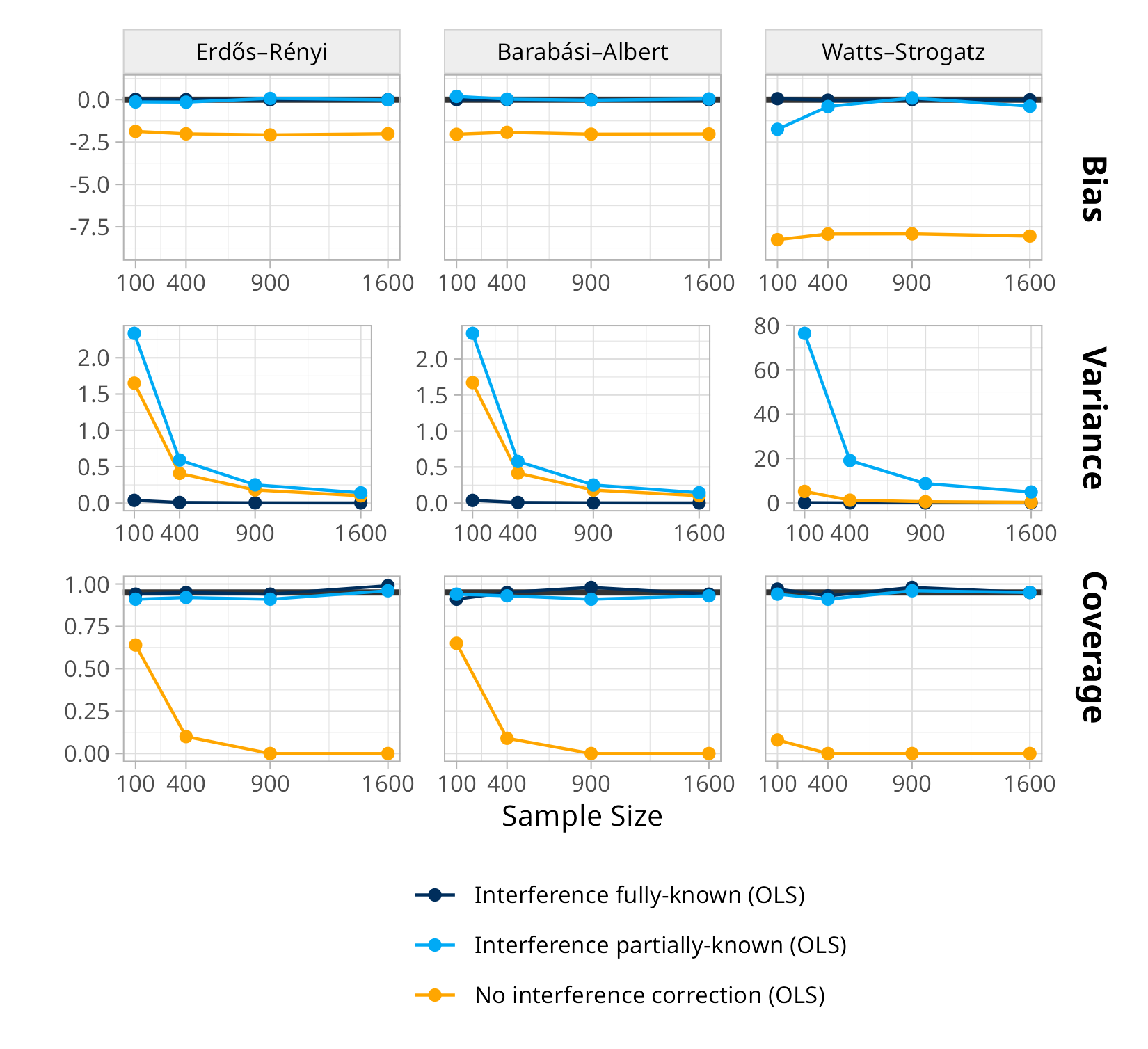}
    \caption{Operating characteristics of homoscedastic linear models under interference.}
    \label{fig:results-homo}
\end{figure}

\begin{figure}[ht]
    \centering
    \includegraphics[width=0.75\textwidth]{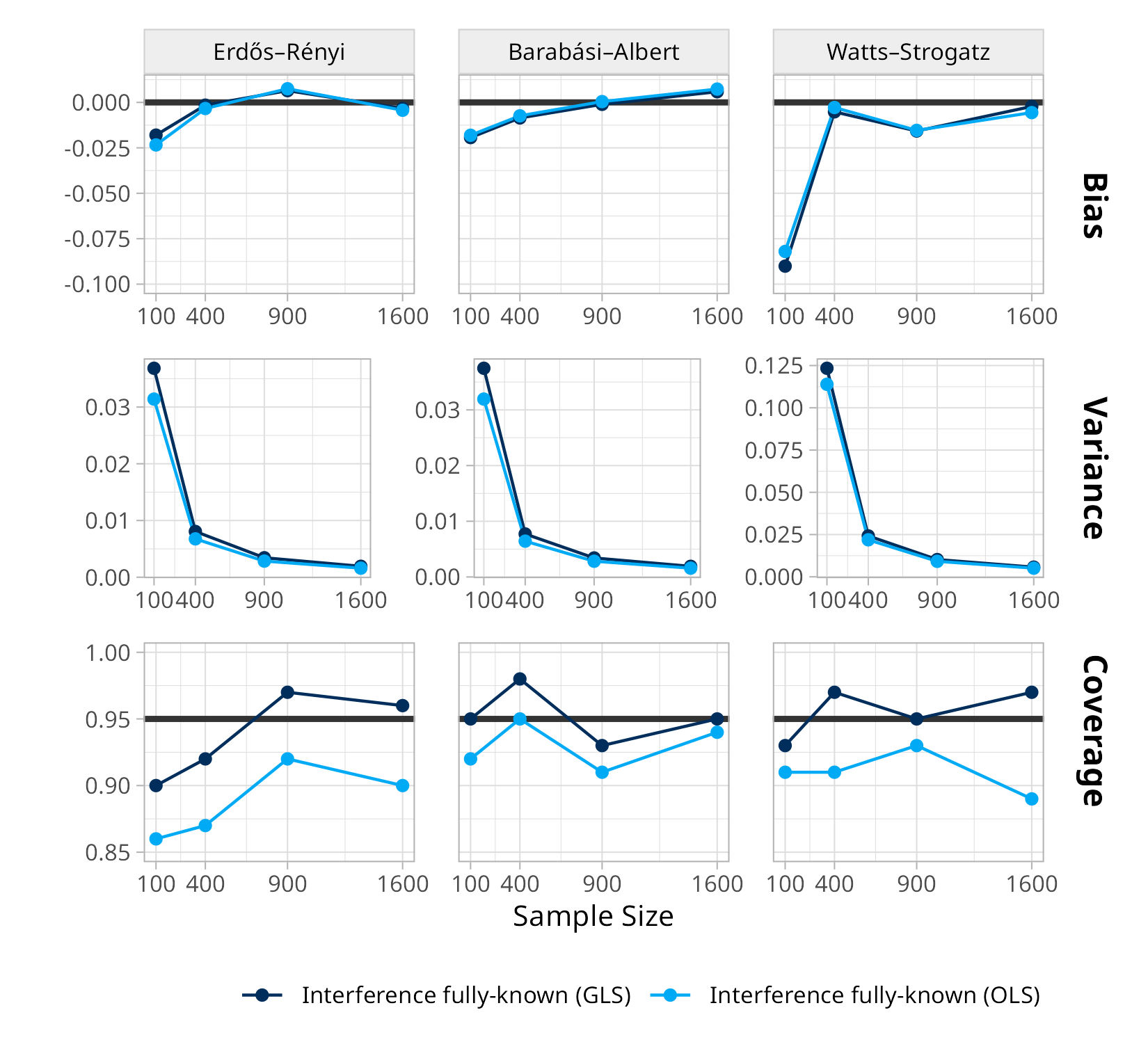}
    \caption{Operating characteristics of correlated-error linear models under interference.}
    \label{fig:results-corr}
\end{figure}

In the homoscedastic setting we compared three estimators: totally-known interference (Section~\ref{sec:totally-known}), partially-known interference (Section~\ref{sec:partially-known}), and a misspecified estimator ignoring interference. In the correlated setting we compared two totally-known estimators: one accounting for correlated errors and one ignoring them. Results are shown in Figures~\ref{fig:results-homo} and~\ref{fig:results-corr}.

As expected, the misspecified model yields biased estimates with poor coverage. Both interference-aware estimators achieve unbiased estimates and approximately nominal 95\% coverage, with variance shrinking with $n$; the variance bias in the Erd\H{o}s-R\'enyi setting was negligible in practice. The totally-known estimator attained substantially lower variance than the partially-known estimator by exploiting full network structure. In the correlated setting, the estimator accounting for error correlation maintained near-nominal coverage, while the one ignoring it did not.

\section{Example Data Analysis}\label{sec:data-analysis}

We illustrate the impact of interference correction on a regression of county-level poverty rates on PM\textsubscript{2.5} across $n = 58$ California counties---a small-sample spatial setting with known low but meaningful effect sizes \citep{Verbeek2021}. County-level PM\textsubscript{2.5} data were obtained from \cite{HDPulse}, while poverty rates and confounders (population density, education, age, race, and share of workers in production or transportation) were extracted from the American Community Survey via \texttt{tidycensus} \citep{census_data}. The interference graph was constructed from LEHD Origin-Destination Employment Statistics \citep{lodes_data}, which tracks home-work commuting patterns and has been used as a mobility proxy in air pollution research \citep{deSouza2023}; edge weights represent the fraction of each county's population commuting into each other county. Summary statistics are given in Table~\ref{tbl:summary}.

\begin{table}[!htbp] \centering
  \caption{Summary statistics of data used in PM\textsubscript{2.5}-poverty regression analysis.}
  \label{tbl:summary}
\begin{tabular}{@{\extracolsep{5pt}}lcccc}
\\[-1.8ex]\hline
\hline \\[-1.8ex]
Statistic & \multicolumn{1}{c}{Mean} & \multicolumn{1}{c}{St. Dev.} & \multicolumn{1}{c}{Min} & \multicolumn{1}{c}{Max} \\
\hline \\[-1.8ex]
Poverty rate & 9.0 & 3.4 & 3.8 & 18.1 \\
PM\textsubscript{2.5} ($\mu$g/m\textsuperscript{3}) & 13.9 & 5.0 & 7.9 & 39.1 \\
Population density (/km\textsuperscript{2}) & 272.1 & 947.2 & 0.7 & 7,030.7 \\
\% in production or transportation jobs & 4.9 & 1.6 & 2.1 & 9.2 \\
Education (\% bachelor's degree or higher) & 29.2 & 12.3 & 12.1 & 61.1 \\
Median age & 40.1 & 6.2 & 31.5 & 54.9 \\
\% White & 62.9 & 15.7 & 34.0 & 89.2 \\
\% Hispanic/Latino & 19.6 & 11.8 & 3.1 & 53.9 \\
Sum of interference weights & 0.2 & 0.1 & 0.1 & 0.5 \\
\hline \\[-1.8ex]
\end{tabular}
\end{table}

We fit two models. First, we fit the interference-aware model
\begin{align*}
    \text{Poverty} \sim\, 1 + \text{PM\textsubscript{2.5}} + \text{density} + \text{prod. jobs} + \text{edu.} + \text{age} + \text{\% white} + \text{\% hisp./latino} \nonumber\\
    + G^\top(1 + \text{PM\textsubscript{2.5}} + \text{density} + \text{prod. jobs} + \text{edu.} + \text{age} + \text{\% white} + \text{\% hisp./latino})
\end{align*}
and applied the formulas from Section~\ref{sec:totally-known} with HC5 heteroscedasticity-consistent variance estimates \citep{CribariNeto2007} from the \texttt{sandwich} package \citep{Zeileis2004} to estimate the MTE and confidence interval. For comparison, we also fit a misspecified model dropping all $G^\top(\cdot)$ terms, taking its PM\textsubscript{2.5} coefficient directly as the MTE.

\begin{figure}[htbp]
    \centering
    \includegraphics[width=0.65\textwidth]{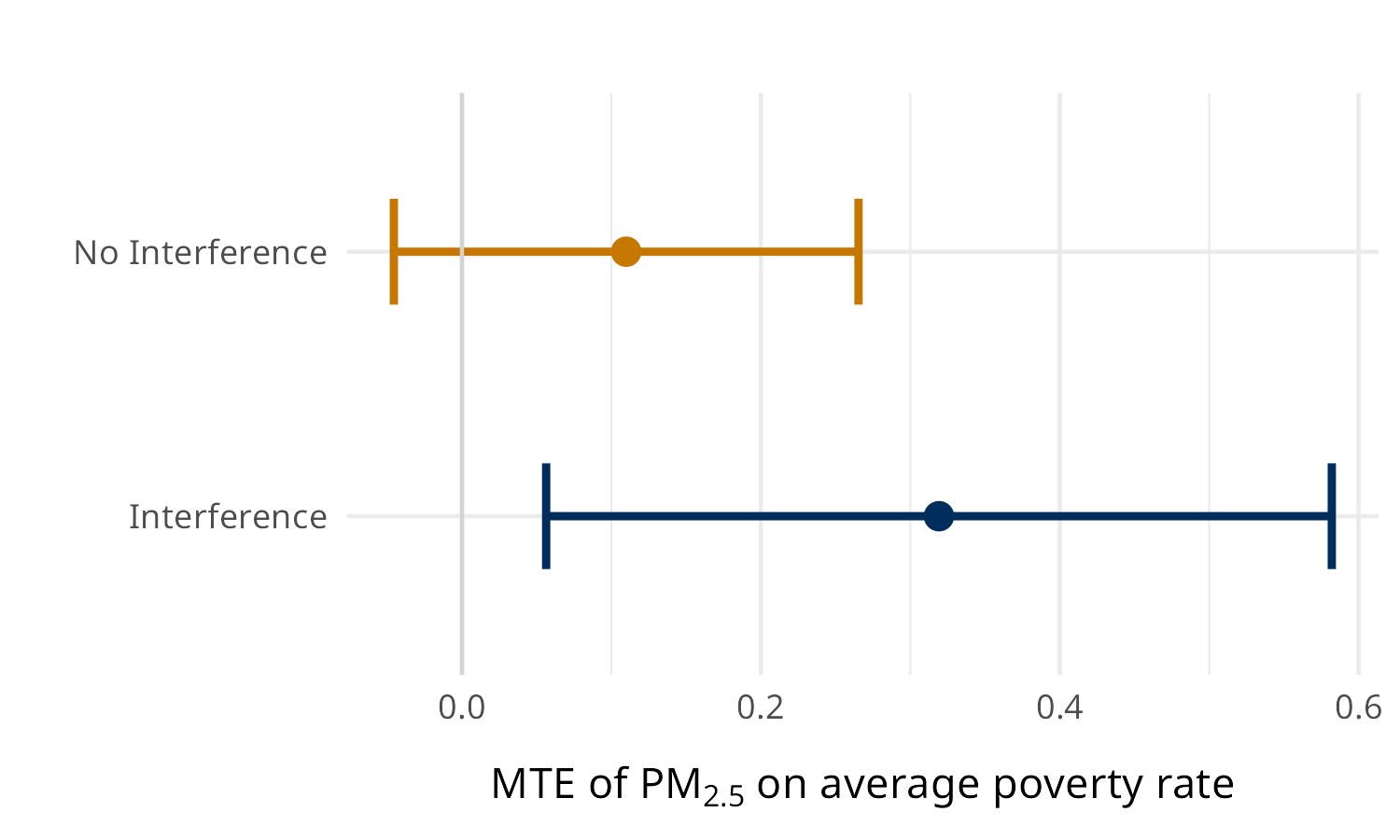}
    \caption{Estimated increase in average poverty rate per 1~$\mu$g/m\textsuperscript{3} increase in PM\textsubscript{2.5}, with and without interference correction.}
    \label{fig:data-analysis}
\end{figure}

Ignoring interference, the estimated MTE of a 1-$\mu$g/m\textsuperscript{3} increase in PM\textsubscript{2.5} on poverty was 0.11 percentage points (95\% CI: $-$0.05, 0.27). Accounting for interference tripled the estimate to 0.32 percentage points (95\% CI: 0.06, 0.58); see Figure~\ref{fig:data-analysis}. An analysis ignoring interference would have yielded an inconclusive result, while accounting for interference reveals a small but statistically significant effect.

\section{Discussion and Conclusions}\label{sec:discussion}

We have shown that under exchangeability and linearity, network interference reduces to a model misspecification problem: the neighbor-summed treatment and covariate terms $A_i^s$ and $L_i^s$ are simply omitted regressors, and correcting for them requires only including them as covariates in a standard linear regression. The resulting plug-in estimators for the total effect are unbiased and consistent, with variance estimates that are exact for fixed networks and asymptotically valid for well-behaved random ones. Extensions to partially known networks, multiple interference structures, correlated errors, and time-varying networks follow naturally from our framework.

The methods as presented apply to main-terms linear models, but extend straightforwardly to models with $A$-$L$ interactions. By simply adding coefficients for interaction terms (scaled by the mean of the covariate being interacted) to the causal identification result, unbiased and consistent estimates for $\psi$ can be constructed using the same strategies as presented in Section \ref{sec:methods}. However, extending these results to generalized linear models with nonlinear link functions is not straightforward: our identification arguments rely on the ability to marginalize over confounder coefficient uncertainty, which fails under nonlinearity for the same reason that the odds ratio is non-collapsible \citep{Whitcomb2020}. To circumvent this problem, one could instead treat the parametric regressoin as a predictive model and applying nonparametric techniques for interference \citep{Balkus2025}, but doing so sacrifices the efficiency and interpretability of the parametric approach. Developing native GLM corrections, perhaps via the delta method applied to maximum likelihood estimates, is a priority for future work, especially for count outcomes where log- or logit-link models are typically preferable to log-transforming the response directly \citep{OHara2010}.

Other directions for future work include extension of our framework to quasi-experimental designs such as difference-in-differences \citep{Xu2023} or regression discontinuity \citep{Torrione2024}, where the identifying assumptions are weaker than those required for the observational ATE and MTE, as well as to scenarios with unknown interference networks \citep{hoshino2023causal, ohnishi2022degree} or networks that may depend on $L$ or $A$. Despite their limitations, the linear models we present offer practitioners a simple, fast, and interpretable entry point for causal inference under interference. Just as ordinary least squares fostered a broad ecosystem of extensions in the independent-units setting, we hope this framework provides a foundation for richer developments in the interference setting.

\section{Supplementary Materials}

Code and data to reproduce the simulation studies and data analysis can be found at \if1\anon{\url{https://github.com/salbalkus/pub-code-linear-model-interference}}\fi 
\if0\anon{\url{https://anonymous.4open.science/r/pub-code-linear-model-interference-2B6F/README.md}}\fi. 

%Raw data extracts are also available at the Harvard Dataverse: \url{https://dataverse.harvard.edu/dataset.xhtml?persistentId=doi:10.7910/DVN/9LMRKZ}.

\section{Acknowledgments}
The authors thank \if1\anon{the National Science Foundation (award no.~DGE 2140743) and the Harvard College and Harvard Data Science Initiative's SPUDS program for financially supporting this work, as well as Sophie Woodward }\fi
\if0\anon{\blackout{the National Science Foundation (award no.~DGE 2140743) and the Harvard College and Harvard Data Science Initiative's SPUDS program for financially supporting this work, as well as Sophie Woodward }}\fi
for helpful feedback during the project's initiation. 

\textit{Conflicts of Interest}: The authors report there are no competing interests to declare.

\textit{Generative AI Statement}: During  research, the authors acknowledge use of GPT4.1 for minimal coding assistance, as well as Gemini 3 and Claude Sonnet 4.6 for stylistic line-editing and typesetting improvements to the written manuscript.

\bibliography{refs}

@article{Naimi2016,
  title = {An Introduction to G Methods},
  ISSN = {1464-3685},
  DOI = {10.1093/ije/dyw323},
  journal = {International Journal of Epidemiology},
  publisher = {Oxford University Press (OUP)},
  author = {Naimi,  Ashley I and Cole,  Stephen R and Kennedy,  Edward H},
  year = {2016},
  month = dec,
  pages = {dyw323}
}

@article{Robins1986,
  title = {A new approach to causal inference in mortality studies with a sustained exposure period—application to control of the healthy worker survivor effect},
  volume = {7},
  ISSN = {0270-0255},
  DOI = {10.1016/0270-0255(86)90088-6},
  number = {9–12},
  journal = {Mathematical Modelling},
  publisher = {Elsevier BV},
  author = {Robins,  James},
  year = {1986},
  pages = {1393–1512}
}

@BOOK{hernan2024,
  TITLE = {Causal Inference: What If. },
  AUTHOR = {Miguel Hern\'{a}n and James Robins},
  YEAR = {2024},
  PUBLISHER = {Chapman \& Hall/CRC.},
  ADDRESS = {Boca Raton, Florida}
}

@article{munozPopulationInterventionCausal2012,
  title = {Population Intervention Causal Effects Based on Stochastic Interventions},
  author = {D{\'i}az, Iv{\'a}n and {van der Laan}, Mark},
  year = {2012},
  journal = {Biometrics},
  volume = {68},
  number = {2},
  pages = {541--549},
  issn = {0006-341X},
  doi = {10.1111/j.1541-0420.2011.01685.x},
}

@article{haneuseEstimationEffectInterventions2013,
  title = {Estimation of the Effect of Interventions That Modify the Received Treatment},
  author = {Haneuse, S. and Rotnitzky, A.},
  year = {2013},
  journal = {Statistics in Medicine},
  volume = {32},
  number = {30},
  pages = {5260--5277},
  issn = {1097-0258},
  doi = {10.1002/sim.5907},
  langid = {english},
  pmid = {23913589}
}

@article{rubin1980,
 ISSN = {01621459},
 author = {Donald B. Rubin},
 journal = {Journal of the American Statistical Association},
 number = {371},
 pages = {591--593},
 publisher = {[American Statistical Association, Taylor & Francis, Ltd.]},
 title = {Randomization Analysis of Experimental Data: The Fisher Randomization Test Comment},
 volume = {75},
 year = {1980}
}

@Manual{r,
     title = {R: A Language and Environment for Statistical Computing},
     author = {{R Core Team}},
     organization = {R Foundation for Statistical Computing},
     address = {Vienna, Austria},
     year = {2025},
     url = {https://www.R-project.org/},
   }

@article{Wu2020,
  title = {Air pollution and COVID-19 mortality in the United States: Strengths and limitations of an ecological regression analysis},
  volume = {6},
  ISSN = {2375-2548},
  DOI = {10.1126/sciadv.abd4049},
  number = {45},
  journal = {Science Advances},
  publisher = {American Association for the Advancement of Science (AAAS)},
  author = {Wu,  X. and Nethery,  R. C. and Sabath,  M. B. and Braun,  D. and Dominici,  F.},
  year = {2020},
  month = nov 
}

@article{Shin2023,
  doi = {10.48550/ARXIV.2305.14194},
  author = {Shin,  Heejun and Braun,  Danielle and Irene,  Kezia and Audirac,  Michelle and Antonelli,  Joseph},
  title = {A spatial interference approach to account for mobility in air pollution studies with multivariate continuous treatments},
  journal = {arXiv:2305.14194},
  year = {2023},
  copyright = {Creative Commons Attribution 4.0 International}
}

@incollection{kempton1997interference,
  title={Interference between plots},
  author={Kempton, RA},
  booktitle={Statistical methods for plant variety evaluation},
  pages={101--116},
  year={1997},
  publisher={Springer}
}

@article{hong2006evaluating,
  title={Evaluating kindergarten retention policy: A case study of causal inference for multilevel observational data},
  author={Hong, Guanglei and Raudenbush, Stephen W},
  journal={Journal of the American Statistical Association},
  volume={101},
  number={475},
  pages={901--910},
  year={2006},
  publisher={Taylor \& Francis}
}

@article{sobel2006randomized,
  title={What do randomized studies of housing mobility demonstrate? {Causal} inference in the face of interference},
  author={Sobel, Michael E},
  journal={Journal of the American Statistical Association},
  volume={101},
  number={476},
  pages={1398--1407},
  year={2006},
  publisher={Taylor \& Francis}
}

@article{Hudgens2008,
  title = {Toward Causal Inference With Interference},
  volume = {103},
  ISSN = {1537-274X},
  DOI = {10.1198/016214508000000292},
  number = {482},
  journal = {Journal of the American Statistical Association},
  publisher = {Informa UK Limited},
  author = {Hudgens,  Michael G and Halloran,  M. Elizabeth},
  year = {2008},
  month = jun,
  pages = {832–842}
}

@article{Ogburn2022,
  title = {Causal Inference for Social Network Data},
  volume = {119},
  ISSN = {1537-274X},
  DOI = {10.1080/01621459.2022.2131557},
  number = {545},
  journal = {Journal of the American Statistical Association},
  publisher = {Informa UK Limited},
  author = {Ogburn,  Elizabeth L. and Sofrygin,  Oleg and Díaz,  Iván and van der Laan,  Mark J.},
  year = {2022},
  month = dec,
  pages = {597–611}
}

@article{hoshino2023causal,
  title={Causal inference with noncompliance and unknown interference},
  author={Hoshino, Tadao and Yanagi, Takahide},
  journal={Journal of the American Statistical Association},
  pages={1--12},
  year={2023},
  publisher={Taylor \& Francis}
}

@article{ohnishi2022degree,
  title={Degree of interference: A general framework for causal inference under interference},
  author={Ohnishi, Yuki and Karmakar, Bikram and Sabbaghi, Arman},
  journal={Journal of Machine Learning Research},
  volume={26},
  number={120},
  pages={1--37},
  year={2025}
}

@book{gelman2007data,
  title={Data analysis using regression and multilevel/hierarchical models},
  author={Gelman, Andrew and Hill, Jennifer},
  year={2007},
  publisher={Cambridge University Press}
}

@article{Fritz2024,
  title={A regression framework for studying relationships among attributes under network interference},
  author={Fritz, Cornelius and Schweinberger, Michael and Bhadra, Subhankar and Hunter, David R},
  journal={Journal of the American Statistical Association},
  number={just-accepted},
  pages={1--22},
  year={2025},
  publisher={Taylor \& Francis}
}

@article{Balkus2025,
  title = {The causal effects of modified treatment policies under network interference},
  ISSN = {1467-9868},
  DOI = {10.1093/jrsssb/qkag052},
  journal = {Journal of the Royal Statistical Society Series B: Statistical Methodology},
  publisher = {Oxford University Press (OUP)},
  author = {Balkus,  Salvador V and Delaney,  Scott W and Hejazi,  Nima S},
  year = {2026},
  month = mar 
}

@article{Emmenegger2025,
  title = {Treatment effect estimation with observational network data using machine learning},
  volume = {13},
  ISSN = {2193-3685},
  DOI = {10.1515/jci-2023-0082},
  number = {1},
  journal = {Journal of Causal Inference},
  publisher = {Walter de Gruyter GmbH},
  author = {Emmenegger,  Corinne and Spohn,  Meta-Lina and Elmer,  Timon and B\"{u}hlmann,  Peter},
  year = {2025},
  month = jan 
}

@article{Tchetgen2010,
  title = {On causal inference in the presence of interference},
  volume = {21},
  ISSN = {1477-0334},
  DOI = {10.1177/0962280210386779},
  number = {1},
  journal = {Statistical Methods in Medical Research},
  publisher = {SAGE Publications},
  author = {Tchetgen,  Eric J Tchetgen and VanderWeele,  Tyler J},
  year = {2010},
  month = nov,
  pages = {55–75}
}

@article{vanderlaanCausalInferencePopulation2014,
  title = {Causal Inference for a Population of Causally Connected Units},
  author = {{van der Laan}, Mark J.},
  year = {2014},
  journal = {Journal of Causal Inference},
  volume = {2},
  number = {1},
  pages = {13--74},
  issn = {2193-3677},
  doi = {10.1515/jci-2013-0002},
  pmcid = {PMC4500386},
  pmid = {26180755}
}

@article{sofryginSemiParametricEstimationInference2017,
  title = {Semi-Parametric Estimation and Inference for the Mean Outcome of the Single Time-Point Intervention in a Causally Connected Population},
  author = {Sofrygin, Oleg and {van der Laan}, Mark J.},
  year = {2017},
  journal = {Journal of Causal Inference},
  volume = {5},
  number = {1},
  pages = {20160003},
  issn = {2193-3677},
  pmcid = {PMC5650205},
  pmid = {29057197},
}

@article{zivichTargetedMaximumLikelihood2022,
  title = {Targeted Maximum Likelihood Estimation of Causal Effects with Interference: A Simulation Study},
  shorttitle = {Targeted Maximum Likelihood Estimation of Causal Effects with Interference},
  author = {Zivich, Paul N. and Hudgens, Michael G. and Brookhart, Maurice A. and Moody, James and Weber, David J. and Aiello, Allison E.},
  year = {2022},
  journal = {Statistics in Medicine},
  volume = {41},
  number = {23},
  pages = {4554--4577},
  issn = {1097-0258},
  doi = {10.1002/sim.9525},
  langid = {english},
  pmcid = {PMC9489667},
  pmid = {35852017}
}

@article{VerbitskySavitz2012,
  title = {Causal Inference Under Interference in Spatial Settings: A Case Study Evaluating Community Policing Program in Chicago},
  volume = {1},
  ISSN = {2161-962X},
  DOI = {10.1515/2161-962x.1020},
  number = {1},
  journal = {Epidemiologic Methods},
  publisher = {Walter de Gruyter GmbH},
  author = {Verbitsky-Savitz,  Natalya and Raudenbush,  Stephen W.},
  year = {2012},
  month = jan 
}

@book{cox_planning_1958,
    title = {Planning of experiments},
    language = {eng},
    publisher = {New York, Wiley},
    author = {Cox, D. R.},
    collaborator = {{Internet Archive}},
    year = {1958},
}

@article{aronow_estimating_2017,
    title = {Estimating average causal effects under general interference, with application to a social network experiment},
    volume = {11},
    issn = {1932-6157, 1941-7330},
    doi = {10.1214/16-AOAS1005},
    number = {4},
    urldate = {2023-05-25},
    journal = {The Annals of Applied Statistics},
    author = {Aronow, Peter M. and Samii, Cyrus},
    month = dec,
    year = {2017},
    note = {Publisher: Institute of Mathematical Statistics},
    pages = {1912--1947},
}

@article{albert2002statistical,
  title={Statistical mechanics of complex networks},
  author={Albert, R{\'e}ka and Barab{\'a}si, Albert-L{\'a}szl{\'o}},
  journal={Reviews of modern physics},
  volume={74},
  number={1},
  pages={47},
  year={2002},
  publisher={APS}
}

@article{watts1998collective,
  title={Collective dynamics of ‘small-world’networks},
  author={Watts, Duncan J and Strogatz, Steven H},
  journal={nature},
  volume={393},
  number={6684},
  pages={440--442},
  year={1998},
  publisher={Nature Publishing Group}
}

@article{deSouza2023,
  title = {Quantifying Disparities in Air Pollution Exposures across the {United States} Using Home and Work Addresses},
  volume = {58},
  ISSN = {1520-5851},
  DOI = {10.1021/acs.est.3c07926},
  number = {1},
  journal = {Environmental Science and Technology},
  publisher = {American Chemical Society (ACS)},
  author = {de Souza,  Priyanka and Anenberg,  Susan and Makarewicz,  Carrie and Shirgaokar,  Manish and Duarte,  Fabio and Ratti,  Carlo and Durant,  John L. and Kinney,  Patrick L. and Niemeier,  Deb},
  year = {2023},
  month = dec,
  pages = {280–290}
}

@inproceedings{Akaike1973,
    title = {Information theory and an extension of the maximum likelihood principle},
    author = {Akaike, Hirotogu},
    booktitle = {2nd international symposium on information theory},
    year = {1973},
    pages = {267-281},
    address = {Budapest, Hungary},
    publisher = {Akad\'{e}mia Kiad\'{o}},
    editor = {N. Petrov and F. Cs\'{a}ki}
}

@article{Rabinowicz2020,
  title = {Cross-Validation for Correlated Data},
  volume = {117},
  ISSN = {1537-274X},
  DOI = {10.1080/01621459.2020.1801451},
  number = {538},
  journal = {Journal of the American Statistical Association},
  publisher = {Informa UK Limited},
  author = {Rabinowicz,  Assaf and Rosset,  Saharon},
  year = {2020},
  month = sep,
  pages = {718–731}
}

@article{Tibshirani1996,
  title = {Regression Shrinkage and Selection Via the Lasso},
  volume = {58},
  ISSN = {1467-9868},
  DOI = {10.1111/j.2517-6161.1996.tb02080.x},
  number = {1},
  journal = {Journal of the Royal Statistical Society Series B: Statistical Methodology},
  publisher = {Oxford University Press (OUP)},
  author = {Tibshirani,  Robert},
  year = {1996},
  month = jan,
  pages = {267–288}
}

@article{Lee2016,
  title = {Exact post-selection inference,  with application to the lasso},
  volume = {44},
  ISSN = {0090-5364},
  DOI = {10.1214/15-aos1371},
  number = {3},
  journal = {The Annals of Statistics},
  publisher = {Institute of Mathematical Statistics},
  author = {Lee,  Jason D. and Sun,  Dennis L. and Sun,  Yuekai and Taylor,  Jonathan E.},
  year = {2016},
  month = jun 
}

@article{OHara2010,
  title = {Do not log‐transform count data},
  volume = {1},
  ISSN = {2041-210X},
  DOI = {10.1111/j.2041-210x.2010.00021.x},
  number = {2},
  journal = {Methods in Ecology and Evolution},
  publisher = {Wiley},
  author = {O’Hara,  Robert B. and Kotze,  D. Johan},
  year = {2010},
  month = may,
  pages = {118–122}
}

@misc{HDPulse,
    title = {{HDPulse: An Ecosystem of Minority Health and Health Disparities Resources}},
    author = {{National Institute on Minority Health and Health Disparities}}, 
    year = {2025},
    url = {\url{https://hdpulse.nimhd.nih.gov}}
}

@Manual{census_data,
  title = {tidycensus: Load US Census Boundary and Attribute Data as 'tidyverse' and 'sf'-Ready Data Frames},
  author = {Kyle Walker and Matt Herman},
  year = {2024},
  note = {R package version 1.6.3},
  url = {https://walker-data.com/tidycensus/},
}

@misc{lodes_data,
  author = {{U.S. Census Bureau}},
  year = {2024},
  publisher = {U.S. Census Bureau, Longitudinal-Employer Household Dynamics Program},
  title = {{LEHD} Origin-Destination Employment Statistics Data (2002-2021)},
  howpublished = {\url{https://lehd.ces.census.gov/data/#lodes}}
}

@article{Lee2024,
  title = {Childhood PM
            2.5
            exposure and upward mobility in the United States},
  volume = {121},
  ISSN = {1091-6490},
  DOI = {10.1073/pnas.2401882121},
  number = {38},
  journal = {Proceedings of the National Academy of Sciences},
  publisher = {Proceedings of the National Academy of Sciences},
  author = {Lee,  Sophie-An Kingsbury and Merlo,  Luca and Dominici,  Francesca},
  year = {2024},
  month = sep 
}

@article{CribariNeto2007,
  title = {Inference Under Heteroskedasticity and Leveraged Data},
  volume = {36},
  ISSN = {1532-415X},
  DOI = {10.1080/03610920601126589},
  number = {10},
  journal = {Communications in Statistics - Theory and Methods},
  publisher = {Informa UK Limited},
  author = {Cribari-Neto,  Francisco and Souza,  Tatiene C. and Vasconcellos,  Klaus L. P.},
  year = {2007},
  month = aug,
  pages = {1877–1888}
}

@article{Zeileis2004,
  title = {Econometric Computing with {HC} and {HAC} Covariance Matrix Estimators},
  volume = {11},
  ISSN = {1548-7660},
  DOI = {10.18637/jss.v011.i10},
  number = {10},
  journal = {Journal of Statistical Software},
  publisher = {Foundation for Open Access Statistic},
  author = {Zeileis,  Achim},
  year = {2004}
}

@article{Whitcomb2020,
  title = {Defining,  Quantifying,  and Interpreting “Noncollapsibility” in Epidemiologic Studies of Measures of “Effect”},
  volume = {190},
  ISSN = {1476-6256},
  DOI = {10.1093/aje/kwaa267},
  number = {5},
  journal = {American Journal of Epidemiology},
  publisher = {Oxford University Press (OUP)},
  author = {Whitcomb,  Brian W and Naimi,  Ashley I},
  year = {2020},
  month = dec,
  pages = {697–700}
}

@article{Bhadra2025,
  author = {Bhadra,  Subhankar and Schweinberger,  Michael},
  title = {Causal Inference Under Network Interference},
  journal = {arXiv:2508.06808},
  year = {2025},
  copyright = {Creative Commons Attribution 4.0 International}
}

@article{tchetgentchetgenAutoGComputationCausalEffects2021,
  title = {{Auto-G-Computation} of causal effects on a network},
  author = {{Tchetgen Tchetgen}, Eric J and Fulcher, Isabel R and Shpitser,
    Ilya},
  year = {2021},
  journal = {Journal of the American Statistical Association},
  volume = {116},
  number = {534},
  pages = {833--844},
  publisher = {Taylor \& Francis},
  issn = {0162-1459},
  doi = {10.1080/01621459.2020.1811098},
  pmid = {34366505},
}

@article{tchetgen2025autodrestimation,
  title = {Auto-doubly robust estimation of causal effects on a network},
  author = {Liu, Jizhou and Zhang, Dake and {Tchetgen Tchetgen}, Eric J},
  year = {2025},
  journal = {arXiv:2506.23332}
}

@article{clarkCausalInferenceStochastic2024,
  title = {Causal inference over stochastic networks},
  volume = {187},
  ISSN = {1467-985X},
  DOI = {10.1093/jrsssa/qnae001},
  number = {3},
  journal = {Journal of the Royal Statistical Society Series A: Statistics in Society},
  publisher = {Oxford University Press (OUP)},
  author = {Clark,  Duncan A and Handcock,  Mark S},
  year = {2024},
  pages = {772–795}
}

@article{Aronow2012,
  title = {A General Method for Detecting Interference Between Units in Randomized Experiments},
  volume = {41},
  ISSN = {1552-8294},
  DOI = {10.1177/0049124112437535},
  number = {1},
  journal = {Sociological Methods \& Research},
  publisher = {SAGE Publications},
  author = {Aronow,  Peter M.},
  year = {2012},
  month = feb,
  pages = {3–16}
}

@article{VazquezBare2023,
  title = {Identification and estimation of spillover effects in randomized experiments},
  volume = {237},
  ISSN = {0304-4076},
  DOI = {10.1016/j.jeconom.2021.10.014},
  number = {1},
  journal = {Journal of Econometrics},
  publisher = {Elsevier BV},
  author = {Vazquez-Bare,  Gonzalo},
  year = {2023},
  month = nov,
  pages = {105237}
}

@article{Torrione2024,
  doi = {10.48550/ARXIV.2410.02727},
  author = {Torrione,  Elena Dal and Arduini,  Tiziano and Forastiere,  Laura},
  title = {Regression Discontinuity Designs Under Interference},
  journal = {arXiv:2410.02727},
  year = {2024},
  copyright = {arXiv.org perpetual,  non-exclusive license}
}

@article{Sinclair2012,
  title = {Detecting Spillover Effects: Design and Analysis of Multilevel Experiments},
  volume = {56},
  ISSN = {1540-5907},
  DOI = {10.1111/j.1540-5907.2012.00592.x},
  number = {4},
  journal = {American Journal of Political Science},
  publisher = {Wiley},
  author = {Sinclair,  Betsy and McConnell,  Margaret and Green,  Donald P.},
  year = {2012},
  month = may,
  pages = {1055–1069}
}

@article{Jiang2022,
  title = {Statistical Inference and Power Analysis for Direct and Spillover Effects in Two-Stage Randomized Experiments},
  volume = {79},
  ISSN = {1541-0420},
  DOI = {10.1111/biom.13782},
  number = {3},
  journal = {Biometrics},
  publisher = {Oxford University Press (OUP)},
  author = {Jiang,  Zhichao and Imai,  Kosuke and Malani,  Anup},
  year = {2022},
  month = oct,
  pages = {2370–2381}
}

@article{Xu2023,
  doi = {10.48550/ARXIV.2306.12003},
  author = {Xu,  Ruonan},
  title = {Difference-in-Differences with Interference},
  journal = {arXiv:2306.12003},
  year = {2023},
  copyright = {Creative Commons Attribution Non Commercial No Derivatives 4.0 International}
}

@article{Baird2018,
  title = {Optimal Design of Experiments in the Presence of Interference},
  volume = {100},
  ISSN = {1530-9142},
  DOI = {10.1162/rest_a_00716},
  number = {5},
  journal = {The Review of Economics and Statistics},
  publisher = {MIT Press},
  author = {Baird,  Sarah and Bohren,  J. Aislinn and McIntosh,  Craig and \"{O}zler,  Berk},
  year = {2018},
  month = dec,
  pages = {844–860}
}

@article{Forastiere2020,
  title = {Identification and Estimation of Treatment and Interference Effects in Observational Studies on Networks},
  volume = {116},
  ISSN = {1537-274X},
  DOI = {10.1080/01621459.2020.1768100},
  number = {534},
  journal = {Journal of the American Statistical Association},
  publisher = {Informa UK Limited},
  author = {Forastiere,  Laura and Airoldi,  Edoardo M. and Mealli,  Fabrizia},
  year = {2020},
  month = jun,
  pages = {901–918}
}

@Manual{igraph,
  title = {{igraph}: Network Analysis and Visualization in R},
  author = {Gábor Csárdi and Tamás Nepusz and Vincent Traag and Szabolcs Horvát and Fabio Zanini and Daniel Noom and Kirill Müller and David Schoch and Maëlle Salmon},
  year = {2025},
  note = {R package version 2.2.1.9013},
  doi = {10.5281/zenodo.7682609},
  url = {https://CRAN.R-project.org/package=igraph},
}

@article{Verbeek2021,
  title = {Systematic Reviews Should Consider Effects From Both the Population and the Individual Perspective},
  volume = {111},
  ISSN = {1541-0048},
  DOI = {10.2105/ajph.2020.306147},
  number = {5},
  journal = {American Journal of Public Health},
  publisher = {American Public Health Association},
  author = {Verbeek,  Jos and Hoving,  Jan and Boschman,  Julitta and Chong,  Lee-Yee and Livingstone-Banks,  Jonathan and Bero,  Lisa},
  year = {2021},
  month = may,
  pages = {820–825}
}

@article{Lbke2020,
  title = {Why We Should Teach Causal Inference: Examples in Linear Regression With Simulated Data},
  volume = {28},
  ISSN = {1069-1898},
  DOI = {10.1080/10691898.2020.1752859},
  number = {2},
  journal = {Journal of Statistics Education},
  publisher = {Informa UK Limited},
  author = {L\"{u}bke,  Karsten and Gehrke,  Matthias and Horst,  J\"{o}rg and Szepannek,  Gero},
  year = {2020},
  month = may,
  pages = {133–139}
}

@book{van2011targeted,
  title={Targeted learning: causal inference for observational and experimental data},
  author={{van der Laan}, Mark J and Rose, Sherri and others},
  volume={4},
  year={2011},
  publisher={Springer}
}

@article{LeSage2010,
  title = {The importance of modeling spatial spillovers in public choice analysis},
  volume = {150},
  ISSN = {1573-7101},
  DOI = {10.1007/s11127-010-9714-6},
  number = {3–4},
  journal = {Public Choice},
  publisher = {Springer Science and Business Media LLC},
  author = {LeSage,  James P. and Dominguez,  Matthew},
  year = {2010},
  month = sep,
  pages = {525–545}
}

@book{anselin1988spatial,
  title={Spatial econometrics: methods and models},
  author={Anselin, Luc},
  volume={4},
  year={1988},
  publisher={Springer Science I\& Business Media}
}

\end{document}